\documentclass[english,aps,manuscript]{revtex4-1}
\usepackage[T1]{fontenc}
\usepackage[latin9]{inputenc}
\usepackage[active]{srcltx}
\usepackage{amssymb}
\usepackage{esint}

\makeatletter
\@ifundefined{textcolor}{}
{%
 \definecolor{BLACK}{gray}{0}
 \definecolor{WHITE}{gray}{1}
 \definecolor{RED}{rgb}{1,0,0}
 \definecolor{GREEN}{rgb}{0,1,0}
 \definecolor{BLUE}{rgb}{0,0,1}
 \definecolor{CYAN}{cmyk}{1,0,0,0}
 \definecolor{MAGENTA}{cmyk}{0,1,0,0}
 \definecolor{YELLOW}{cmyk}{0,0,1,0}
}

\makeatother

\usepackage{babel}
\begin{document}

\title{Entropy Spectrum of a Carged Black Hole of Heterotic String Theory
via Adiabatic Invariance}

\author{Alexis Larrañaga}

\email{ealarranaga@unal.edu.co}

\selectlanguage{english}%

\address{National Astronomical Observatory. National University of Colombia.}

\author{Luis Cabarique}

\email{lfcabariqueb@unal.edu.co}

\selectlanguage{english}%

\address{Department of Physics. National University of Colombia.}

\author{Manuel Londo\~{n}o}

\email{malondonoc@unal.edu.co}

\selectlanguage{english}%

\address{Department of Physics. National University of Colombia.}

\begin{abstract}
Using adiabatic invariance and the Bohr-Sommerfeld quantization rule
we investigate the entropy spectroscopy of a charged black hole of
heterotic string theory. It is shown that the entropy spectrum is
equally spaced identically to the Schwarzschild, Reissner-Nordström
and Kerr black holes. Since the adiabatic invariance method does not
use quasinormal mode analysis, there is no need to impose the small
charge limit and no confusion on whether the real part or imaginary
part is responsible for the entropy spectrum. 


\end{abstract}
\maketitle

\section{Introduction}

Bekenstein \cite{Bekenstein} showed that the black hole horizon area
is an adiabatic invariant, so that it should be quantized by Ehrenfest
principle. Considering the minimum change of the horizon area, $\triangle A$,
due to the absorption of a test particle into the black hole, Bekenstein
argued that it can only be proportional to $\hbar$, so that the area
spectrum should be linearly quantized. This equally spaced area spectrum
let us understand the entropy of a black hole as proportional to the
horizon area because it can be considered as formed by patches of
equal Planck's area $\ell_{p}^{2}$ and the patches get added one
at a time. Using the point particle model presented in \cite{Christodoulou1970},
Bekenstein found that the smallest possible increase in horizon area
of a non-extremal black hole is exactly $\Delta A=8\pi\ell_{p}^{2}$. 

Recent investigation by Hod \cite{Hod1998} and later by Maggiore
\cite{Maggiore2008} on quasinormal modes has shown the possibility
to use Bohr's correspondence principle to provide a method to quantize
the horizon area of black holes. On the other hand, Zeng et. al. \cite{Zeng2012}
obtined the area spectrum of Schwarzschild and Kerr black holes by
considering that the frequency of an outgoing wave is given by the
inverse of the Hawking temperature.

In this work, we will investigate the spectroscopy of a charged black
hole of heterotic string theory by the new approach presented by Mahji
and Vagenas \cite{Majhi}. Here the entropy spectrum is obtained by
using solely the adiabaticity of black holes and the Bohr-Sommerfeld
quantization rule. The resulting black hole entropy shows an equally
spaced spectrum and the corresponding horizon area spectrum is identical
to the result reported by Wei et. al. \cite{Wei2010} using quasinormal
modes.

\section{The GMGHS black hole}

The low energy effective action of the heterotic string theory in
four dimensions is given by

\begin{equation}
\mathcal{A}=\int d^{4}x\sqrt{-\tilde{g}}e^{-\psi}\left(-R+\frac{1}{12}H_{\mu\nu\rho}H^{\mu\nu\rho}-\tilde{G}^{\mu\nu}\partial_{\mu}\psi\partial_{\nu}\psi+\frac{1}{8}F_{\mu\nu}F^{\mu\nu}\right),
\end{equation}

where $R$ is the Ricci scalar, $\tilde{G}_{\mu\nu}$ is the metric
that arises naturally in the $\sigma$ model and

\begin{equation}
F_{\mu\nu}=\partial_{\mu}A_{\nu}-\partial_{\nu}A_{\mu}
\end{equation}
is the Maxwell field associated with a $U\left(1\right)$ subgroup
of $E_{8}\times E_{8}$. There is also a dilaton field $\psi$ and
an antisymmetric tensor gauge field $B_{\mu\nu}$ given by

\begin{equation}
H_{\mu\nu\rho}=\partial_{\mu}B_{\nu\rho}+\partial_{\nu}B_{\rho\mu}+\partial_{\rho}B_{\mu\nu}-\left[\Omega_{3}\left(A\right)\right]_{\mu\nu\rho},
\end{equation}
with the gauge Chern-Simons term

\begin{equation}
\left[\Omega_{3}\left(A\right)\right]_{\mu\nu\rho}=\frac{1}{4}\left(A_{\mu}F_{\nu\rho}+A_{\nu}F_{\rho\mu}+A_{\rho}F_{\mu\nu}\right).
\end{equation}
Taking $H_{\mu\nu\rho}=0$ and working in the conformal Einstein frame,
the action becomes

\begin{equation}
\mathcal{A}=\int d^{4}x\sqrt{-\tilde{g}}\left(-R+2\left(\nabla\psi\right)^{2}+e^{-2\phi}F^{2}\right),
\end{equation}
where the Einstein frame metric $\tilde{g}_{\mu\nu}$ is defined as

\begin{equation}
\tilde{g}_{\mu\nu}=e^{-\psi}\tilde{G}_{\mu\nu}.
\end{equation}

The charged black hole solution to the corresponding field equations
is known as the Gibbons-Maeda-Garfinkle-Horowitz-Strominger (GMGHS)
solution \cite{gmghs,Wei2010} and can be written as

\begin{eqnarray}
ds^{2} & = & -f\left(r\right)dt^{2}+\frac{dr^{2}}{f\left(r\right)}+r\left(r-\frac{Q^{2}e^{-2\psi_{0}}}{M}\right)d\Omega^{2}\label{eq:GMGHS}\\
e^{-2\psi} & = & e^{-2\psi_{0}}\left(1-\frac{Q^{2}}{Mr}\right)\\
F & = & Q\sin\theta d\theta\wedge d\vartheta
\end{eqnarray}
where $f\left(r\right)=1-\frac{2M}{r}$, $M$ and $Q$ are the mass
and electric charge of the black hole respectively and $\psi_{0}$
is the asymptotic value of the dilaton. There is a spherical event
horizon at

\begin{equation}
r_{H}=2M\label{eq:horizon}
\end{equation}

with an area given by

\begin{eqnarray}
A & = & 4\pi r_{H}^{2}-8\pi Q^{2}e^{-2\psi_{0}}.\label{eq:area}
\end{eqnarray}

From Eq. (\ref{eq:area}) it is easy to see that the GMGHS solution
becomes a naked singularity if

\begin{equation}
r_{H}^{2}\leq2Q^{2}e^{-2\psi_{0}},
\end{equation}
or equivalently when
\begin{equation}
M^{2}\leq\frac{1}{2}Q^{2}e^{-2\psi_{0}}.
\end{equation}

The surface gravity at the horizon takes the value

\begin{center}
\begin{equation}
\kappa=\frac{1}{2}f'(r_{H})=\frac{1}{4M}\label{eq:surfacegravity}
\end{equation}

\par\end{center}

and therefore the Hawking temperature is

\begin{equation}
T=\frac{\hbar\kappa}{2\pi}=\frac{1}{8\pi M}\label{eq:hawkingtemperature}
\end{equation}

which is independent of the electric charge. Finally, the electric
potential at the event horizon is 
\begin{equation}
\phi=\frac{Q}{r_{H}}e^{-2\psi_{0}}
\end{equation}

and the relation between area and entropy gives 

\begin{equation}
S=\frac{A}{4\hbar}=\frac{\pi r_{H}^{2}-2\pi Q^{2}e^{-2\psi_{0}}}{\hbar}.\label{eq:entropy}
\end{equation}

\section{Entropy Spectrum}

Now we will use the adiabatic invariant action 
\begin{equation}
I=\sum_{i}\int p_{i}dq_{i}
\end{equation}
 to obtain the entropy spectrum of the GMGHS black hole. We have

\begin{center}
\begin{equation}
I=\sum_{i}\int\int_{o}^{p_{i}}dp'_{i}dq_{i}=\sum_{i}\int\int_{o}^{H}\frac{dH'}{\dot{q_{i}}}dq_{i}
\end{equation}

\par\end{center}

where $p_{i}$ is the conjugate momentum of the $q_{i}$ coordinate.
Here we take $i=0,1$ with $q_{0}=\tau$ the Euclidean time, $q_{1}=r$
and dot means derivative with respect to $\tau$. Therefore, this
invariant can be expanded as

\begin{center}
\begin{equation}
I=\int\int_{o}^{H}dH'd\tau+\int\int_{o}^{H}\frac{dH'}{\dot{r}}dr.\label{eq:aux1}
\end{equation}

\par\end{center}

Considering the black hole metric (\ref{eq:GMGHS}), it can be ``Euclideanized''
by the transformation $t\rightarrow-i\tau$, giving 

\begin{center}
\begin{equation}
ds^{2}=f(r)d\tau^{2}+\frac{dr^{2}}{f(r)}+r\left(r-\frac{Q^{2}e^{-2\psi_{0}}}{M}\right)d\Omega^{2}.
\end{equation}

\par\end{center}

Considering radial null paths we have 

\begin{center}
\begin{equation}
\dot{r}=\frac{dr}{d\tau}=\pm i\left|f(r)\right|=R_{\pm}(r)
\end{equation}

\par\end{center}

where the $\pm$ sign mean outgoing or ingoing paths respectively.
Using this relation we can write 

\begin{center}
\begin{equation}
\int\int_{o}^{H}dH'd\tau=\int\int_{o}^{H}dH'\frac{dr}{R_{\pm}(r)}=\int\int_{o}^{H}dH'\frac{dr}{\dot{r}}
\end{equation}

\par\end{center}

and the adiabatic invariant (\ref{eq:aux1}) becomes

\begin{center}
\begin{equation}
I=2\int\int_{o}^{H}dH'd\tau.
\end{equation}

\par\end{center}

The Euclidean time $\tau$ is periodic with period $\frac{2\pi}{\kappa}$
where $\kappa$ is the surface gravity given by Eq. (\ref{eq:surfacegravity}).
Now, we will only consider outgoing paths and therefore the limits
in the $\tau$ integration will be $0$ and $\frac{\pi}{\kappa}$,
obtaining

\begin{center}
\begin{equation}
I=2\pi\int_{o}^{H}\frac{dH'}{\kappa}=\hbar\int_{o}^{H}\frac{dH'}{T}.
\end{equation}

\par\end{center}

To relate this result with the entropy of the black hole, consider
the first law of thermodynamics for the GMGHS black hole 

\begin{equation}
dM=TdS+\phi dQ.
\end{equation}

Following the work of Wald \cite{wald93} and Sudarsky \& Wald \cite{sudarsky}
to obtain the Hamiltonian of the Einstein-Maxwell system we have the
relation of $H$ with the mass and electric potential,

\begin{equation}
H=M-\phi Q.
\end{equation}
Therefore, the adiabatic invariant in terms of the entropy is simply

\begin{center}
\begin{equation}
I=\hbar S.
\end{equation}

\par\end{center}

Implementing the Bohr-Sommerfeld quantization rule

\begin{center}
\begin{equation}
I=\sum_{i}\int p_{q_{i}}dq_{i}=2\pi n\hbar
\end{equation}

\par\end{center}

gives the equally spaced entropy spectrum 

\begin{center}
\[
S=2\pi n
\]

\par\end{center}

or

\begin{center}
\[
\triangle S=2\pi.
\]

\par\end{center}

Using the relation between entropy and horizon area (\ref{eq:entropy})
we obtain the area spectrum

\begin{center}
\[
\triangle A=8\pi l_{p}^{2}
\]

\par\end{center}

where $l_{p}$ is Planck's lenght (in units with $G=c=1$).

\section{Conclusions}

Although the quantum gravity theory has not been found, it is meaningful
to investigate quantum corrections to the entropy spectrum. Here we
have investigated the entropy of a GMGHS black hole with the help
of Bohr-Sommerfeld quantization rule and the adiabatic invariance.
The result shows an equally spaced entropy spectrum which is the same
as the one of Schwarzschild, Reissner-Nordström and Kerr black holes.
This fact confi{}rms the proposal of Bekenstein that the area spectrum
of a black hole is independent of its parameters. Evenmore, our result
is identical to the spectrum reported by Wei et. al. \cite{Wei2010}
using the quasinormal modes analysis.

However, the adiabatic invariance calculation does not need the quasinormal
mode frequency, so there is no confusion on whether the real part
or imaginary part is responsible for the entropy spectrum and the
small charge limit, which is necessary in the quasinormal mode analysis,
was not imposed. In a forthcoming paper we will extend this analysis
to the charged rotating black hole of heterotic string theory found
by Sen \cite{Sen}.

\emph{Acknowledgements}

This work was supported by the Universidad Nacional de Colombia. Hermes
Project Code 13038.

\end{document}